# Superconductivity in KCa$_2$Fe$_4$As$_4$F$_2$ with Separate Double Fe$_2$As$_2$ Layers

Zhi-Cheng Wang,[†] Chao-Yang He,[†] Si-Qi Wu,[†] Zhang-Tu Tang,[†] Yi Liu,[†] Abduweli Ablimit,[†] Chun-Mu Feng,[†] and Guang-Han Cao[*,†,‡]

[†] Department of Physics and State Key Lab of Silicon Materials, Zhejiang University, Hangzhou 310027, China;

[‡] Collaborative Innovation Centre of Advanced Microstructures, Nanjing 210093, China

**ABSTRACT:** We report synthesis, crystal structure and physical properties of a quinary iron-arsenide fluoride KCa$_2$Fe$_4$As$_4$F$_2$. The new compound crystallizes in a body-centered tetragonal lattice (with space group $I4/mmm$, $a$ = 3.8684(2) Å, $c$ = 31.007(1) Å, and $Z$ = 2), which contains double Fe$_2$As$_2$ conducting layers separated by insulating Ca$_2$F$_2$ layers. Our measurements of electrical resistivity, dc magnetic susceptibility and heat capacity demonstrate bulk superconductivity at 33 K in KCa$_2$Fe$_4$As$_4$F$_2$.

**TEXT:**

Currently, only cuprate and iron-based superconductors exhibit ambient-pressure high-temperature superconductivity.[1,2] The two classes of materials share many similarities.[3] Structurally, for example, they are quasi-two-dimensional, possessing superconductively active CuO$_2$ planes and Fe$_2$As$_2$ layers, respectively. This feature makes it possible to "design" a new superconductor by replacing the crystallographic spacer layers.[4,5] So far, however, all the Fe-based superconductors (FeSCs) discovered consist of separate Fe$_2$As$_2$ monolayer or infinite Fe$_2$As$_2$ layers [such as the case of (Ba,K)Fe$_2$As$_2$]. By contrast, many cuprate superconductors contain separate multi-CuO$_2$ planes, which may optimize the superconducting transition temperature $T_c$.[3,4]

Based on an understanding of crystal chemistry of FeSCs, especially in terms of "hard and soft acids and bases" (HSAB)[6] concept, we proposed a rational structural design in search for new FeSCs.[5] Nine structures containing Fe$_2X_2$ ($X$ refers to a pnictogen or a chalcogen element) layers were suggested to be the candidates. Among them, the one with a chemical formula of $A_3$Fe$_4X_4Z_2$,[5] where $A$ represents "hard" cations (there are two distinct crystallographic sites for $A$) and $Z$ denotes a "hard" anion, contains separate double Fe$_2$As$_2$ layers. According to the HSAB rule,[6] $AZ$ and Fe$X$ are expected to combine together, respectively, forming fluorite-like and anti-fluorite-like layers. Fig. 1 shows an example of $A_3$Fe$_4X_4Z_2$, KCa$_2$Fe$_4$As$_4$F$_2$, which is exactly the title compound. The crystal structure is actually an intergrowth of ZrCuSiAs-type CaFeAsF and ThCr$_2$Si$_2$-type KFe$_2$As$_2$. Indeed, the crystallographic building unit [KFe$_4$As$_4$]$^{2-}$ (highlighted by the red rectangular block in the middle of Fig. 1) contains double Fe$_2$As$_2$ layers (with K atoms sandwiched) separated by the Ca$_2$F$_2$ layers, precisely resembling the double CuO$_2$ planes in cuprates exemplified by La$_{2-x}$Sr$_x$CaCu$_2$O$_6$.[7] The analogous crystal structure was earlier reported in Cu- and Ni-based pnictides,[8] however, their Fe-based counterpart has never been synthesized until present, to the best of our knowledge.

In this Communication, we report synthesis, crystal structure and superconductivity of $KCa_2Fe_4As_4F_2$. Owing to the intergrowth structure, the $Fe_2As_2$ layers with different cations aside are no longer symmetric. The material is hole doped by itself, which yields a superconducting transition temperature $T_c$ of 33 K.

The $KCa_2Fe_4As_4F_2$ polycrystalline sample was synthesized via solid-state reactions of pre-synthesized/pre-treated KAs, $CaF_2$, CaAs and $Fe_2As$ in an evacuated container. First, the intermediate binary products of KAs, CaAs and $Fe_2As$ were prepared by solid-state reactions of their corresponding elements at 250, 750 and 750 °C, respectively, in evacuated quartz tubes. The source materials are K cubes (99.5%), redistilled Ca granules (99.5%), Fe powders (99.998%), As pieces (99.999%) and $CaF_2$ powders (99.5%). $CaF_2$ was heated to 500 °C for 12 hours in a muffle furnace to remove adsorbed water. Second, stoichiometric mixtures of KAs, $CaF_2$, CaAs and $Fe_2As$ were pressed into pellets, and the pellets were loaded in an alumina tube with a cover. Then the sample-loaded alumina tube was sealed in a Ta tube which was jacketed with an evacuated quartz ampoule. The sample assembly was sintered at 930 °C for 36 hours. Third, the solid-state reaction was repeated, with a sample homogenization by grinding, in order to improve sample's quality. The final product is stable in air.

We employed an energy dispersive x-ray (EDX) spectroscopy to verify the chemical composition of the title compound. The result shows that the crystalline grains are indeed composed of K, Ca, Fe, As and F, whose atomic ratios basically meet the expected chemical formula (see Figure S1 and Table S1 in the Supporting Information). Powder X-ray diffraction (XRD) experiments were conducted at room temperature on a PANalytical X-ray diffractometer with Cu K$\alpha$1 radiation. Based on the structural model shown in Fig. 1, the crystal structure was refined by a Rietveld analysis using RIETAN software.[9] The resulting weighted reliable factor $R_{wp}$ is 4.04% and the goodness-of-fit $\chi^2$ is 1.85, indicating reliability of the refinement.

The electrical resistivity ($\rho$), Hall coefficient ($R_H$) and heat capacity ($C$) measurements were carried out on a Quantum Design physical property measurement system (PPMS-9). We employed the ac transport option for the $\rho(T,H)$ and $R_H(T)$ measurements. The $C(T,H)$ data were measured by a thermal relaxation method using a square-shaped sample plate (13.1 mg). The dc magnetization was measured on a Quantum Design magnetic property measurement system (MPMS-XL5). The sample was polished into a rod whose demagnetization factor is 0.14. Both zero-field cooling (ZFC) and field-cooling (FC) protocols were adopted for probing the superconducting transition.

The XRD pattern of the as-prepared sample can be well indexed with a body-centered tetragonal lattice. Impurity phases of $KFe_2As_2$, FeAs and $CaF_2$ can be identified, but their strongest lines are only 3.0%, 2.0% and 1.8%, respectively, of the strongest reflection of the main phase. We then carried out a three-phase Rietveld refinement (the reflection lines of $CaF_2$ were taken as an internal standard, which was not included for the refinement), as shown in Fig. 2. The result shows that the mass percentage of the main phase $KCa_2Fe_4As_4F_2$ is 93.4%.

Table 1 lists the crystallographic data of $KCa_2Fe_4As_4F_2$ refined. The $a$ axis lies between those of $KFe_2As_2$ (3.842 Å)[10] and CaFeAsF (3.878 Å)[11]. Meanwhile, the $c$ axis is noticeably smaller than, yet very close to, the sum (31.05 Å) of that of $KFe_2As_2$ (13.861 Å)[10] and twice of that of CaFeAsF ($2 \times 8.593$ Å)[11], consistent with the proposed intergrowth "12442" structure in Fig. 1. The shrinkage in $c$ axis suggests stabilization of the intergrowth phase. While the thickness of the "$KFe_2As_2$" block is obviously shortened (from 6.931 Å to 6.871 Å), the length of the "CaFeAsF" unit is elongated appreciably (from 8.593 Å to 8.632 Å). This structural variation is associated with charge redistribution, because the Fe valence turns into 2.25+, instead of being either 2+ in CaFeAsF or 2.5+ in KFe$_2$As$_2$. Similar charge homogenization appears in $AeA$Fe$_4$As$_4$ ($Ae$ = Ca, Sr; $A$ = K, Rb, Cs)[12] and RbEuFe$_4$As$_4$[13] (1144-type) superconductors discovered very recently.

It is the hybridized structure that makes the Fe$_2$As$_2$ layers asymmetric, unlike most FeSCs that possess the Fe$_2$As$_2$ layers with S$_4$ symmetry. Such an asymmetric Fe coordination was first observed in the 1144-type superconductors[12,13] (but these FeSCs still contain infinite Fe$_2$As$_2$ layers). In FeSCs, the As height from the Fe plane and the As–Fe–As bond angle ($\alpha$) are considered to be the two relevant structural parameters that determine $T_c$.[14,15] Here, the As1 (nearby K) and As2 (nearby Ca) heights are 1.405(3) Å and 1.436 (3) Å, respectively, which seem to meet a $T_c$ value of about 30 K. Meanwhile, the bond angles As1–Fe–As1 and As2–Fe–As2 are 108.0(2)° and 106.8(2)°, respectively, both of which are not far from the ideal value of 109.5° for acquiring an optimized $T_c$.

Figure 3 shows the resistivity measurement result for the polycrystalline KCa$_2$Fe$_4$As$_4$F$_2$ sample. The $\rho(T)$ curve exhibits a metallic behavior above $T_c$. A broad hump appears around 175 K, which is very often seen in hole-doped FeSCs.[16-18] This phenomenon can be explained in terms of an incoherent-to-coherent crossover, which is very recently revealed as an emergent Kondo-lattice behavior in heavily hole-doped $Ak$Fe$_2$As$_2$ ($Ak$ = K, Rb, Cs).[18] The hole-doping scenario is unambiguously demonstrated by the Hall measurement which shows positive $R_H(T)$ values above $T_c$ (see Figure S2 in the Supporting Information). Below 100 K, the resistivity decreases almost linearly until superconductivity emerges. The superconducting transition is clearly shown in Fig. 3(b) where the onset, midpoint, and zero-resistance temperatures are determined to be 33.1, 32.8 and 32.3 K, respectively. Note that the temperature at which $\rho(T)$ deviates from linearity is as high as 35 K, indicating a strong superconducting thermal fluctuation. Similar phenomenon is observed in Sr$_2$VFeAsO$_3$ which hosts a very weak interlayer coupling because of the thick spacer layers.[19] Here the weak interlayer coupling seems to be related to the body-centered lattice, because the adjacent Fe$_2$As$_2$ double layers mutually shift by (1/2, 1/2, 0).

Upon applying magnetic fields, the superconducting transitions shift to lower temperatures, as seen in Fig. 3(c). Impressively, the superconducting transition severely broadens with pronounced tails under magnetic fields, similar to the case in Sr$_2$VFeAsO$_3$.[19] Using a conventional criterion of 90% $\rho_n$ ($\rho_n$ refers to the linearly extrapolated normal-state resistivity at $T_c$ under zero field) for determining the upper critical field ($H_{c2}$) and, 1% $\rho_n$ for the irreversible field ($H_{irr}$), we obtain a superconducting phase diagram [Fig. 3(d)]. The slope of $H_{c2}(T)$ is as high as 8.4 T K$^{-1}$, suggesting a small superconducting coherence length at zero temperature. The large gap between $H_{c2}(T)$ and $H_{irr}(T)$ implies a high anisotropy owing to the weak interlayer coupling. Future measurements under very high magnetic fields and using single-crystal samples may give an accurate and expanded superconducting phase diagram.

Superconductivity in KCa$_2$Fe$_4$As$_4$F$_2$ is confirmed by the dc magnetic susceptibility ($\chi$) measurement. As shown in Fig. 4, the onset diamagnetic transition temperature is 33 K, consistent with the resistivity measurement above. The upper limit of the superconducting volume fraction, measured in ZFC process, achieves 100% at temperatures far below $T_c$, after making a demagnetization correction. On the other hand, the volume fraction of magnetic repulsion, measured in FC process, is remarkably reduced owing to a magnetic flux pinning effect which is clearly demonstrated by the isothermal magnetization at 2 K (shown in the left inset of Fig. 4). Even so, the magnetic repulsion fraction (11%), a lower limit of the superconducting volume fraction, is still higher than any impurity content in the sample (in fact, none of the identified impurities are possibly superconducting at 33 K). Therefore, $KCa_2Fe_4As_4F_2$ is unambiguously responsible for the 33-K superconductivity. Note that there is a detectable kink at about 4 K, close to the superconducting transition of $KFe_2As_2$.[17] The magnitude of the decrease in $4\pi\chi$ (down to 2.0 K) is about 2%, satisfying the $KFe_2As_2$ content from the XRD result above.

Figure 5 shows the specific heat capacity for $KCa_2Fe_4As_4F_2$. An obvious specific-heat jump at $T_c$ = 33 K can be seen in the inset of Fig. 5(a), which further confirms bulk superconductivity. The thermodynamic transition temperature is 32.7 K, based on an entropy-conserving construction. Under a magnetic field of 8 T, the specific-heat jump is remarkably suppressed [Fig. 5(b)]. As a result, the difference in $C(T)$ exhibits a peak at around $T_c$. The thermodynamic transition temperature is 0.5 K lowered at 8 T. If utilizing the thermodynamic transitions for determining $H_{c2}$, the initial slope of $H_{c2}(T)$ can be as high as 16 T K$^{-1}$. As mentioned above, measurements on the single crystals will be able to clarify this discrepancy.

The magnitude of the specific-heat jump scaled by $T_c$, achieves $\Delta C/T_c$ = 150 mJ K$^{-2}$ mol$^{-1}$ (or 38 mJ K$^{-2}$ mol-Fe$^{-1}$). In the BCS weak-coupling scenario, the dimensionless parameter $\Delta C/(\gamma T_c)$ equals to 1.43, where $\gamma$ denotes the normal-state electronic specific-heat coefficient. If so, the $\gamma$ value can be estimated to be 105 mJ K$^{-2}$ mol$^{-1}$ (equivalent to 26 mJ K$^{-2}$ mol-Fe$^{-1}$). This large $\gamma$ value explains the high-temperature $C(T)$ data that unusually exceed the Dulong-Petit limit. Noted here is that the enhanced $\gamma$ is not unusual for hole-doped FeSCs.[13,20]

As is known, most FeSCs are realized through chemical doping in so-called parent compounds.[3,5] Here the occurrence of superconductivity in $KCa_2Fe_4As_4F_2$ is also due to hole doping, yet by the material itself. As the constituents of $KCa_2Fe_4As_4F_2$, $CaFeAsF$[11] and $KFe_2As_2$[17] are a parent compound and a 3.8-K superconductor, respectively. The structural hybridization changes the Fe valence into 2.25+, as mentioned above. This means that the title compound is actually hole doped to a level of 0.25 holes/Fe. Indeed, the Hall carrier number is estimated to be 0.18 holes/Fe if assuming a single-band scenario and employing the Hall coefficient value at 100 K (Figure S2). Note that similar self-doping effect, albeit via an internal charge transfer, was found in other systems including $Sr_2VFeAsO_3$,[21] $Ba_2Ti_2Fe_2As_4O$[22] and $Eu_3Bi_2S_4F_4$[23]. One may expect that such a self-doping strategy could be continually utilized for future exploration of superconductors.

Finally, we remark on the prospect of our finding. In the 12442-type superconductor, there are five different crystallographic sites, which are suitably occupied by different type of elements according to the HSAB classification. Hence there are a lot of combinations for the element replacements in $KCa_2Fe_4As_4F_2$. Hopefully, more 12442-type superconductors will be discovered through the simple element replacements in the near future[24].

**AUTHOR INFORMATION**

**Corresponding Authors:** * ghcao@zju.edu.cn

**ACKNOWLEDGMENT**

This work was supported by NSF of China (Grants No. 11190023 and 90922002) and the Fundamental Research Funds for the Central Universities of China.

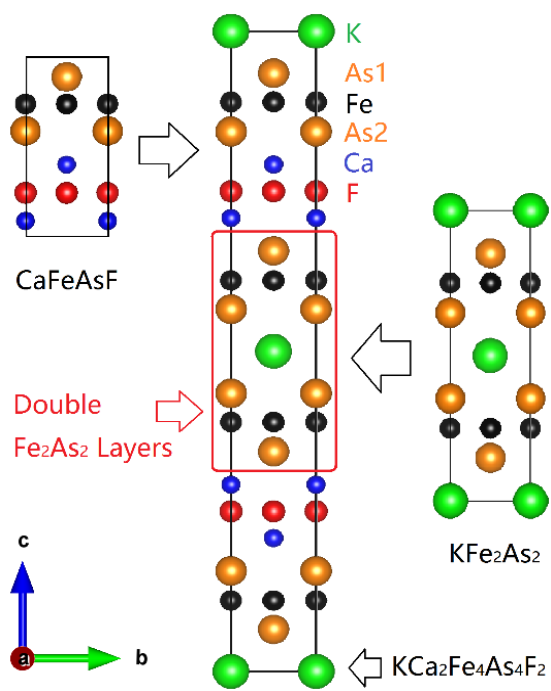

**Figure 1.** Crystals structure of KCa$_2$Fe$_4$As$_4$F$_2$ with separate double Fe$_2$As$_2$ layers (highlighted with the red rectangular block in the middle), resulted from an intergrowth of CaFeAsF (left) and KFe$_2$As$_2$ (right).

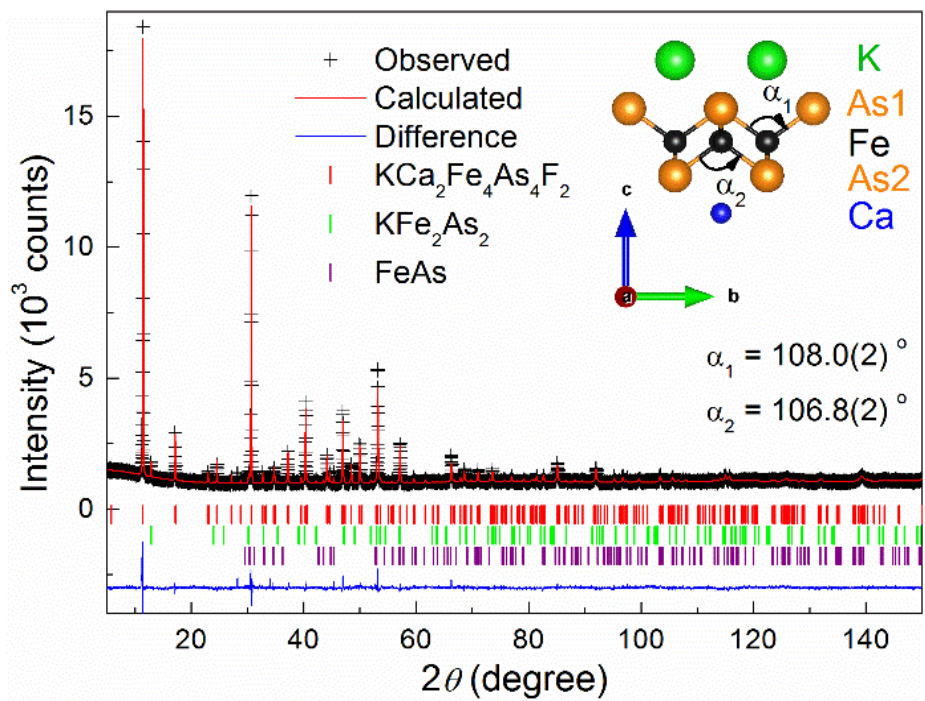

**Figure 2.** Multiple-phase Rietveld refinement profile of powder X-ray diffraction of the KCa$_2$Fe$_4$As$_4$F$_2$ sample. The inset shows structure of the Fe$_2$As$_2$ layer refined, projected along [100] direction.

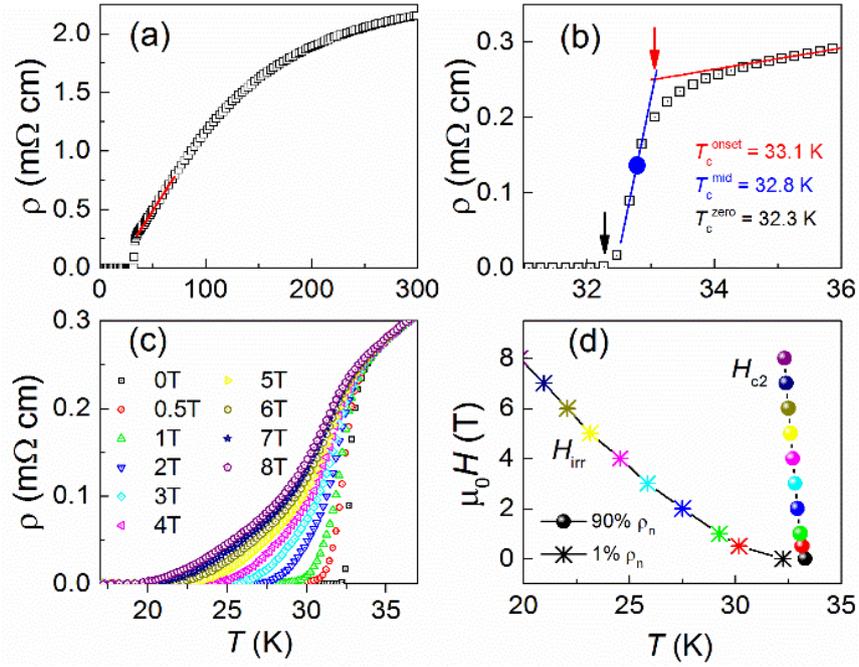

**Figure 3.** (a-c) Temperature dependence of resistivity for the KCa$_2$Fe$_4$As$_4$F$_2$ polycrystalline sample. (d) Derived upper critical field ($H_{c2}$) and irreversible field ($H_{irr}$).

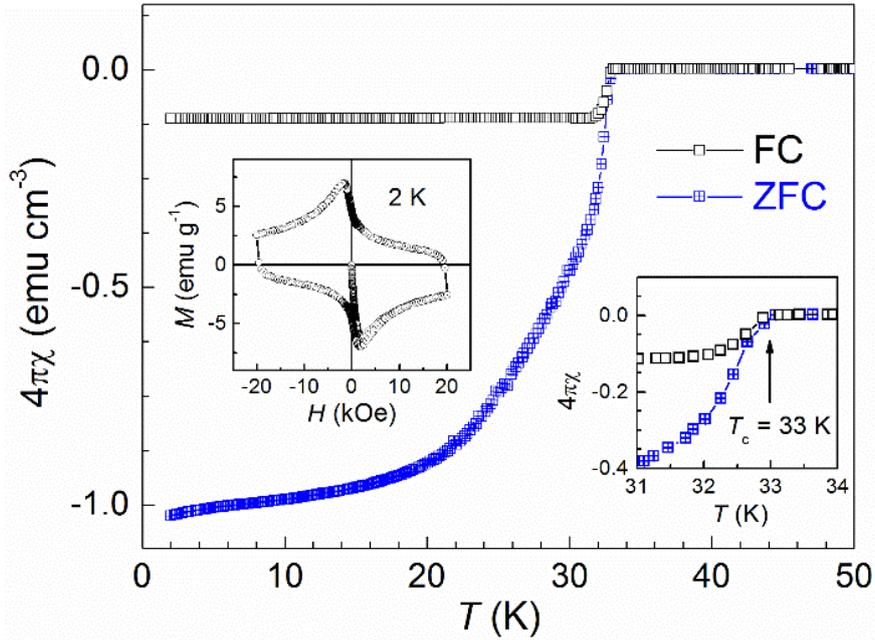

**Figure 4.** Temperature dependence of magnetic susceptibility measured at 10 Oe for KCa$_2$Fe$_4$As$_4$F$_2$. The left inset shows the isothermal magnetization loop at 2 K, and the right-side inset zooms in the superconducting transition.

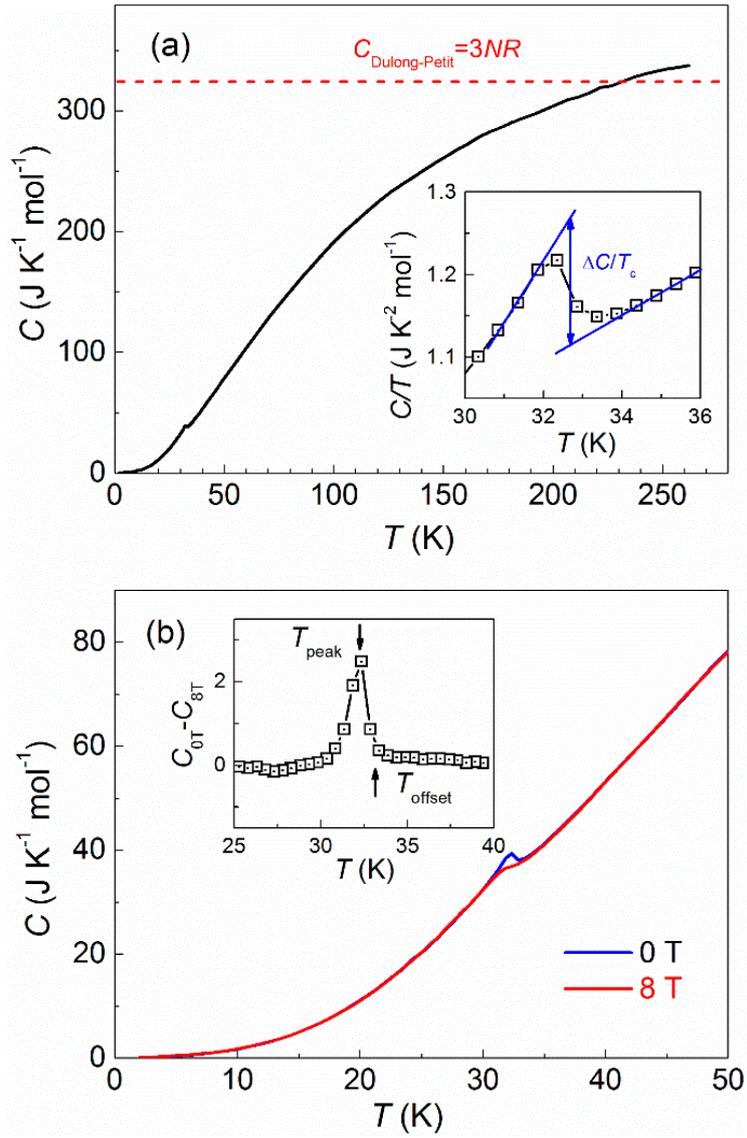

**Figure 5.** (a) Specific heat $C(T)$ in the whole temperature range for $KCa_2Fe_4As_4F_2$. The inset plots $C/T$ in a narrow temperature range, showing a specific-heat jump at $T_c$. (b) Comparison of the $C(T)$ data under magnetic fields of $\mu_0 H = 0$ and 8 T. Their difference is displayed in the inset.

Table 1. Crystallographic data for $KCa_2Fe_4As_4F_2$ (Space Group: *I4/mmm*, No. 139) at room temperature. The lattice parameters are $a$ = 3.8684(2) Å and $c$ = 31.007(1) Å.

| atom | site | x | y | z | Occ.* | $B_{iso}$ # |
|---|---|---|---|---|---|---|
| K | 2a | 0 | 0 | 0 | 1.0 | 1.2 (3) |
| Ca | 4e | 0 | 0 | 0.2085(2) | 1.0 | 0.2 |
| Fe | 8g | 0 | 0.5 | 0.1108(1) | 1.0 | 0.2 |
| As1 | 4e | 0 | 0 | 0.0655(1) | 1.0 | 0.4 (1) |
| As2 | 4e | 0 | 0 | 0.1571(1) | 1.0 | 0.3 (1) |
| F | 4d | 0 | 0.5 | 0.25 | 1.0 | 1.0 |

* The occupancy of each atom is fixed to be 1.0 in the Rietveld refinement. # The temperature factors of Ca, Fe and F are fixed to avoid an unphysical negative value.

********************************************************************************

Supporting Information

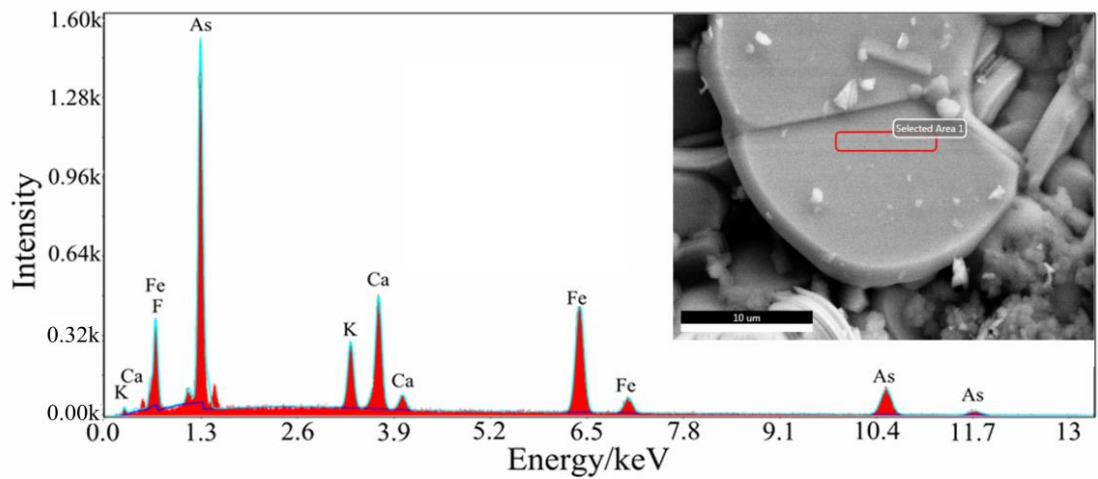

**Figure S1:** A typical energy dispersive X-ray spectrum taken on the fresh surface (shown in the inset) of $KCa_2Fe_4As_4F_2$ polycrystals. The quantitative analysis is given in Table S1.

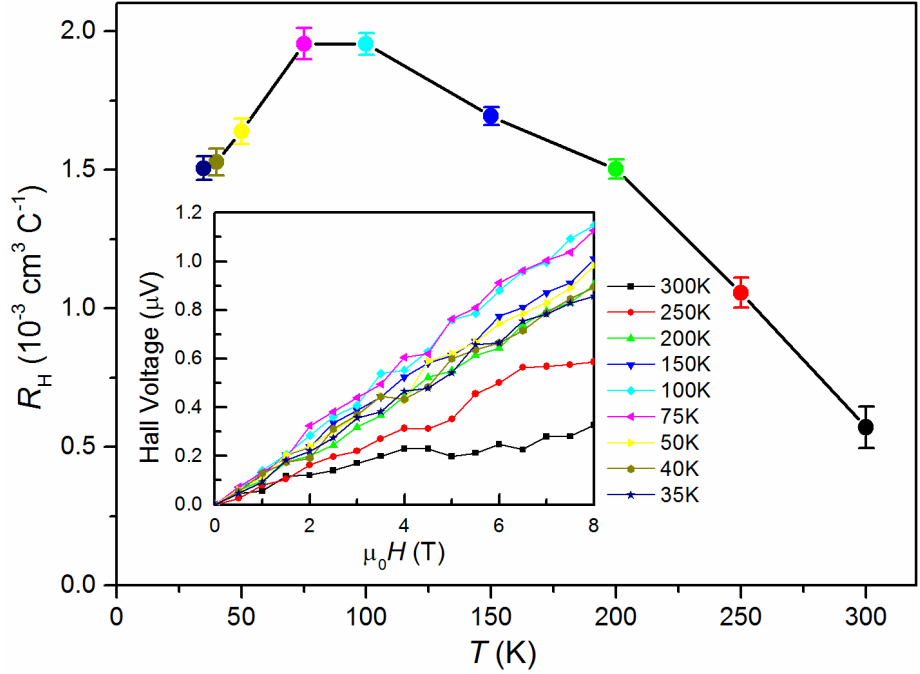

**Figure S2:** Hall coefficient ($R_H$) as a function of temperature measured on KCa$_2$Fe$_4$As$_4$F$_2$ polycrystalline sample. The inset shows field dependence of Hall voltage, from which $R_H$ is obtained by a linear fit. The magnitude of error bars in $R_H(T)$ is taken as three times large as the standard deviations of the linear fit.

**Table S1:** Chemical composition determination of the KCa$_2$Fe$_4$As$_4$F$_2$ sample via energy dispersive x-ray spectroscopy. The atomic ratios are scaled by assuming that the total number of atoms is 13, which gives a chemical formula of K$_{1.0(1)}$Ca$_{2.0(1)}$Fe$_{3.5(2)}$As$_{3.9(3)}$F$_{2.7(3)}$. This formula basically meets the ideal one within the measurement uncertainties. Note that the overlap of the peak of F and Fe in Fig. S1 may lead to a systematic deviation for the two elements.

| Element  Exper. # | K | Ca | Fe | As | F |
|---|---|---|---|---|---|
| 1 | 1.02 | 1.96 | 3.70 | 4.05 | 2.28 |
| 2 | 0.99 | 1.76 | 3.66 | 4.02 | 2.57 |
| 3 | 1.02 | 2.04 | 3.47 | 3.55 | 2.93 |
| 4 | 0.96 | 2.00 | 3.50 | 3.56 | 2.99 |
| 5 | 0.84 | 2.01 | 3.22 | 4.24 | 2.70 |
| Mean Value | 0.96 | 1.95 | 3.51 | 3.88 | 2.69 |
| Standard Deviation | 0.07 | 0.11 | 0.19 | 0.31 | 0.29 |